\documentclass[conference]{IEEEtran}
\usepackage{amsmath}
\usepackage{url}

\usepackage{amssymb}
\usepackage{mdwlist}
\usepackage{fmtcount}

\ifCLASSINFOpdf
   \usepackage[pdftex]{graphicx}
\usepackage{algorithmic}
\usepackage{algorithm}

\usepackage{array}

\usepackage[tight,footnotesize]{subfigure}

\title{Automated Selection of Uniform Regions for CT Image Quality Detection}
\author{Maitham D Naeemi$^{1}$, Adam M Alessio$^{2}$, Sohini Roychowdhury$^{1}$\\
$^{1}$Department of Electrical Engineering, University of Washington, Bothell WA\\
$^{2}$Department of Radiology, University of Washington, Seattle WA\\
}

\begin{document}
%

%


\maketitle

\begin{abstract}

CT images are widely used in pathology detection and follow-up treatment procedures. Accurate identification of pathological features requires diagnostic quality CT images with minimal noise and artifact variation. In this work, a novel Fourier-transform based metric for image quality (IQ) estimation is presented that correlates to additive CT image noise. In the proposed method, two windowed CT image subset regions are analyzed together to identify the extent of variation in the corresponding Fourier-domain spectrum. The two square windows are chosen such that their center pixels coincide and one window is a subset of the other. The Fourier-domain spectral difference between these two sub-sampled windows is then used to isolate spatial regions-of-interest (ROI) with low signal variation (ROI-LV) and high signal variation (ROI-HV), respectively. Finally, the spatial variance ($var$), standard deviation ($std$), coefficient of variance ($cov$) and the fraction of abdominal ROI pixels in ROI-LV ($\nu'(q)$), are analyzed with respect to CT image noise. For the phantom CT images, $var$ and $std$ correlate to CT image noise ($|r|>0.76$ ($p\ll0.001$)), though not as well as $\nu'(q)$ ($r=0.96$ ($p\ll0.001$)). However, for the combined phantom and patient CT images, $var$ and $std$ do not correlate well with CT image noise ($|r|<0.46$ ($p\ll0.001$)) as compared to $\nu'(q)$ ($r=0.95$ ($p\ll0.001$)). Thus, the proposed method and the metric, $\nu'(q)$, can be useful to quantitatively estimate CT image noise.
\end{abstract}


%

\section{Introduction}
Computed Tomography (CT) is a widely used imaging technology that allows volumetric visualization of the internal micro-structure of a scanned object \cite{CT}. Various medical applications can be found in thoracic, cardiac, angiographic and colon imaging where CT images play a crucial role in pathology diagnosis and treatment management. CT image acquisition can operate at various combinations of machine settings, such as the x-ray tube voltage ($kV_p$) and current ($mA$), detector collimation, helical pitch, reconstruction filter, slice thickness, and so on. Associated with each scan setting, the resulting CT volumetric data will have different resolution and noise properties, which can subsequently impact the accuracy of pathology detection \cite{zeng}. Additionally, the present day CT imaging community faces serious concerns regarding the risks of diagnostic radiation exposure. This encourages the implementation of optimized CT imaging protocols with sufficient diagnostic \textit{image quality} (IQ) at the lowest achievable dosage levels. The goal is to maintain contrast enhanced dosages \textit{As Low As Reasonably Achievable (ALARA)} while also ensuring that IQ is sufficient for accurate diagnoses \cite{boone}. In this paper, we propose a novel CT image noise metric that can be used to give automated feedback regarding CT IQ in an effort to attain ALARA. 

Current attempts at evaluating IQ rely on 1-D and 2-D correlations in multi-slice CT scans \cite{zeng}, noise estimation by conventional approaches \cite{mcgibney}, maximum-likelihood \cite{aja}, Bayesian maximum-a-posteriori \cite{aja}, linear estimator \cite{manjon}, or adaptive non-local means estimates \cite{aja1}. However, none of these methods have yet been translated to clinical use.

In this work, we propose a novel CT IQ metric that correlates the degree of spatial variation surrounding each pixel in an image. In uniform density (attenuation) regions, this variation is assumed to be an additive combination of signal and random noise. CT IQ is inversely proportional to the additive noise, such that low noise leads to high quality images and vice versa. Additionally, noise estimation, and therefore CT IQ classification, should be improved by extracting regions of low signal variations. This is achieved by relying on the observation that, in the absence of noise, the Fourier-domain estimate of a uniform spatial region of an image is independent of sample size. Thus, variations in the Fourier-domain spectrum of different sample sizes within a uniform spatial region can be indicative of additive noise, and the degree of variation in the Fourier-domain spectra can be used as a quantitative metric for IQ.

This paper makes two key contributions. First, a novel windowed Fourier-domain based distance metric (WFDM) is introduced that is capable of estimating the variation in a relatively uniform spatial region due to additive noise. WFDM is found to identify thresholded spatial regions that correlate with the image acquisition parameters. Second, the WFDM is analyzed using phantom and real patient CT image data to ensure robustness and reliability of the proposed metric.

\section{Proposed Method}
For quantitative assessment of CT IQ, we extract spatial regions of interest (ROI) that can be used to estimate noise variance. CT images can be assumed to contain additive noise along with the signal component and hence, segmentation of relatively uniform regions in every CT image is imperative for image quality estimation tasks.

\begin{figure}[h!]
	\begin{center}
		\subfigure[]{\includegraphics[width = 1.05in, height=.9in]{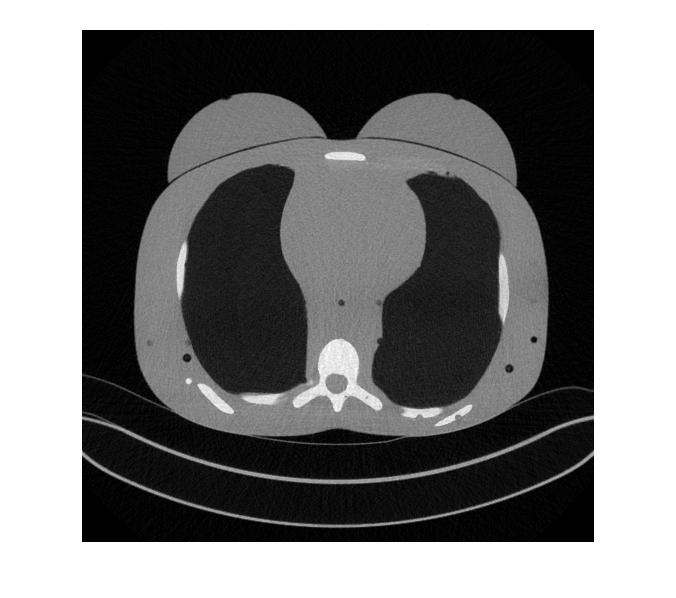}}
		\subfigure[]{\includegraphics[width = 1.05in, height=.9in]{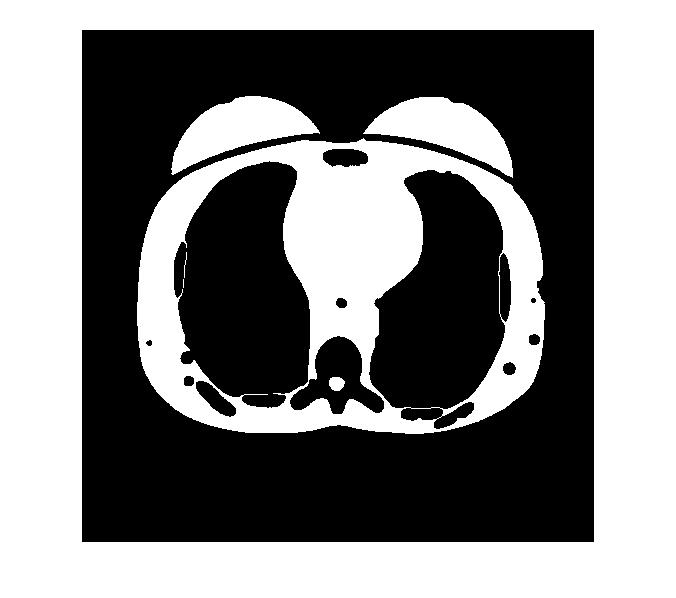}}
		\subfigure[]{\includegraphics[width = 1.05in, height=.9in]{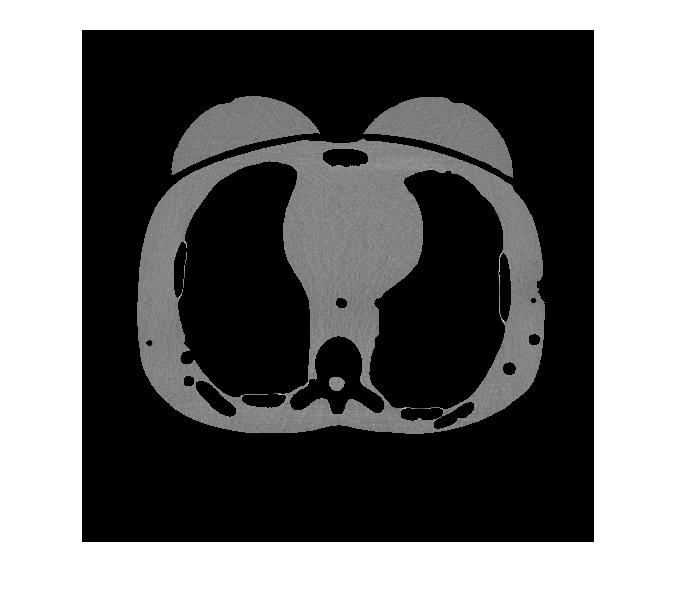}}
		\caption{CT image thresholding for region of interest (ROI) extraction. (a) The original image of a phantom acquired at 10$mA$ tube current. (b) A binary mask of the ROI generated by fixed thresholding after scaling the original pixel values to the range of [0,1] and setting [0.5,0.6] = 1, [0,0.5) \& (0.6,1] = 0. (c) The extracted ROI resulting from pixel-wise multiplication of the binary mask and original image.}     \label{roi}
	\end{center}
\end{figure}

A baseline method of ROI selection in CT images is performed by fixed thresholding, where all pixel values within a given range are set to $1$ while values out of that range are set to $0$. This baseline method is illustrated in Fig. \ref{roi}, where a CT image with pixel values scaled to $[0$,$1]$ is thresholded so that all pixels within $[0.5$,$0.6]$ are set to $1$ and the rest to $0$. The threshold pixel intensity range is empirically determined. The thresholding results in the binary image shown in Fig.\ref{roi}(b) that can serve as a mask to extract a relatively uniform ROI shown in Fig.\ref{roi}(c). However, fixed thresholding selects ROI based on absolute pixel intensity, and hence cannot be generalized to ROI-LV extraction due to variations in pixel intensity across patients and different tissue types.

Prior work has shown that in the Fourier-domain, the central frequency spectrum structure (CFSS) of an image primarily corresponds to the signal rather than noise \cite{roych}. Therefore, the CFSS primarily correlates with the signal, and can be used to compare different regions of the image.  

Specifically, for an abdominal CT image, the ROI-LV region is observed to have a Fourier-domain frequency spectrum that is more compact in the central, low frequency region than the in the higher frequencies. Further, the CFSS of the CT image data under analysis is observed to be mostly independent of the size of ROI-LV. Conversely, the spatial ROI with high variation (ROI-HV) is observed to have a Fourier-domain frequency spectrum with a less compact CFSS than the higher frequency regions and depend on sample size. Thus, changes in the CFSS due to changing sample size can be used to identify ROI-LV within an image or subset of an image.

Therefore, we propose a \textit{windowed Fourier-domain distance metric (WFDM)} as a means of locating regions of low spatial variation that can be used to better correlate noise and CT IQ.

\subsection{The Windowed Fourier-domain Distance Metric (WFDM)}
\begin{figure}[ht]
\begin{center}
\includegraphics[width = 3.2in, height=2in]{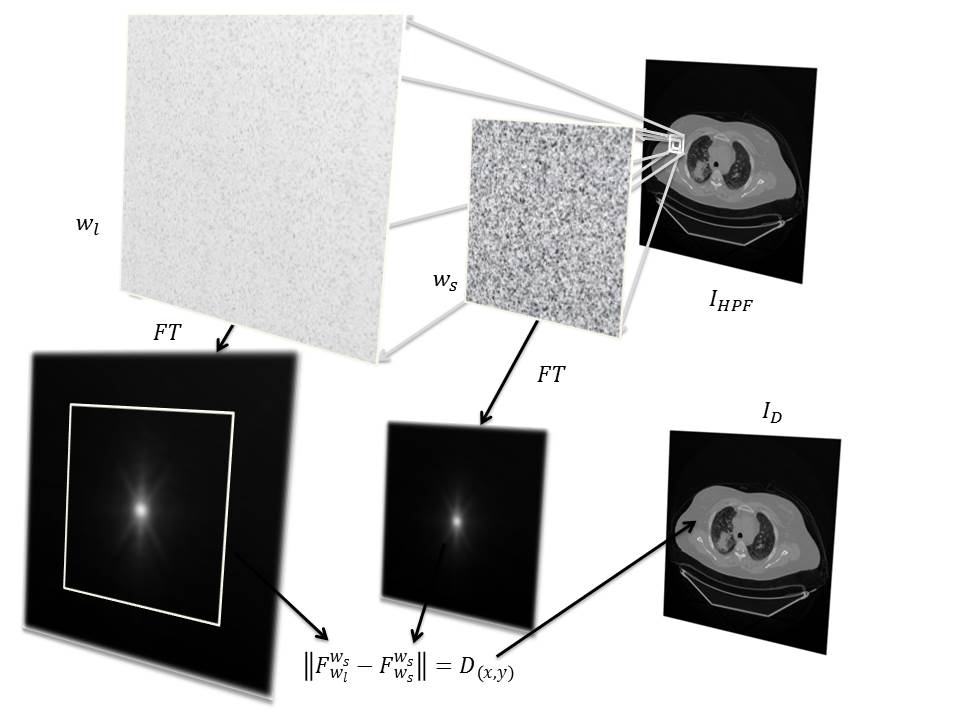}
\caption{The high-pass filtered image, $I_{HPF}$, is analyzed in the Fourier-domain of two subsets of different size, $w_l$ and $w_s$. The Euclidean distance between the two spectra, $D_{(x,y)}$, is found and stored at position $(x,y)$ in the distance image, $I_D$.}     \label{scheme}
\end{center}
\end{figure}

The WFDM process follows the schematic shown in Fig. \ref{scheme} and detailed below.

We begin with a noisy image, $I_n$, of size $[n\times n]$  (\ref{img}) where $I$ is the ideal image and $N$ is the additive noise. This image is passed through a high-pass filter (\ref{hpf}), so that the local average intensities do not bias the measurement of low or high local spatial variation.
\begin{align}
I_n &= I+N \label{img} \\
I_{HPF} &= HPF(I_n). \label{hpf}
\end{align}

At this point, two square windows centered at $(x,y)$ are used to extract subsets of the image. The smaller window, $w_s$, has a dimension of $[d_s\times d_s]$ (\ref{ws}) while the larger window, $w_l$, has a dimension of $[d_l\times d_l]$ (\ref{wl}). 
\begin{align}
\forall (x,y), \quad w_s &= I_{HPF}(x,y)[d_s\times d_s], \label{ws}\\
w_l &= I_{HPF}(x,y)[d_l\times d_l], \quad d_l>d_s. \label{wl}
\end{align}

The Fourier-transform of these spatially windowed regions are computed as $F^{w_s}$ and $F^{w_l}$ in (\ref{fws}) and (\ref{fwl}), respectively.
\begin{align}
F^{w_s} &= \mathcal{F}(w_s), \label{fws}\\
F^{w_l} &= \mathcal{F}(w_l), \label{fwl}\\
F^{w_s}_{w_s} &= F^{w_s}[w_s], \label{fwsws}\\
F^{w_l}_{w_s} &= F^{w_l}[w_s]. \label{fwlws}
\end{align}

A region of size $w_s$ around the center of each spectrum, ($F^{w_s},F^{w_l}$), is sub-sampled, in (\ref{fwsws}) and (\ref{fwlws}) respectively. The Euclidean distance (\ref{dist}) between the resulting spectra, ($F_{w_s}^{w_l},F_{w_s}^{w_s}$), is then found, corresponding to changes in the CFSS between the two spectra. We previously observed that the CFSS of ROI-LV is independent of sample size, as opposed to that of ROI-HV. Thus, this distance, $D_{(x,y)}$, is a Fourier-domain based measure of spatial uniformity surrounding each pixel $(x,y)$, and is stored in its respective location in a \textit{distance image}, $I_D$ (\ref{id}).
\begin{align}
D_{(x,y)} &= \|F_{w_s}^{w_l}-F_{w_s}^{w_s}\|_2, \label{dist}\\
I_D(x,y) &= D_{(x,y)}. \label{id}
\end{align}

The intensity of each pixel in $I_D$ correlates the degree of spatial variation surrounding that pixel in the image $I_n$, hence regions with similar variations in pixel strength can be segmented by fixed thresholding $I_D$ to find ROI-LV and ROI-HV, respectively.

Similar to the fixed thresholding used in the baseline approach shown in Fig.\ref{roi}, a threshold $q$ can be applied to $I_D$ to generate a mask, $U^q_T$ (\ref{ut}). This mask can then be used to segment the original image $I_n$ by regions of similar spatial variation and thus provides a generalizable approach to ROI-LV selection.
\begin{align}
U^q_T(x,y) = \begin{cases} 1, & \mbox{if } I_D(x,y) \leq q\\ 0, & \mbox{if }I_D(x,y) > q\end{cases} \label{ut}
\end{align}

\subsection{Data}
The data used for this work came from both phantom and patient scans. From an IRB approved study, CT images were acquired on two patients with a Lightspeed 16-slice scanner, where most settings were kept constant (helical, pitch 1.375, 20$mm$ collimation, 120$kVp$) and only the tube current was varied. For the patient whose scans were used in this study, the tube current values were 48$mA$ and 5$mA$, respectively. Both patient studies had 354 image slices from head to toe, but we extracted only the 101 image slices that correspond to the abdominal CT region. A dataset of phantom CT images was acquired using the tube current values of $10,25,75,125,175,$ and $350mA$, respectively.

\subsection{CT Image Visualization}
\begin{figure}[h!]
\begin{center}
\subfigure[]{\includegraphics[width = 1in, height=0.8in]{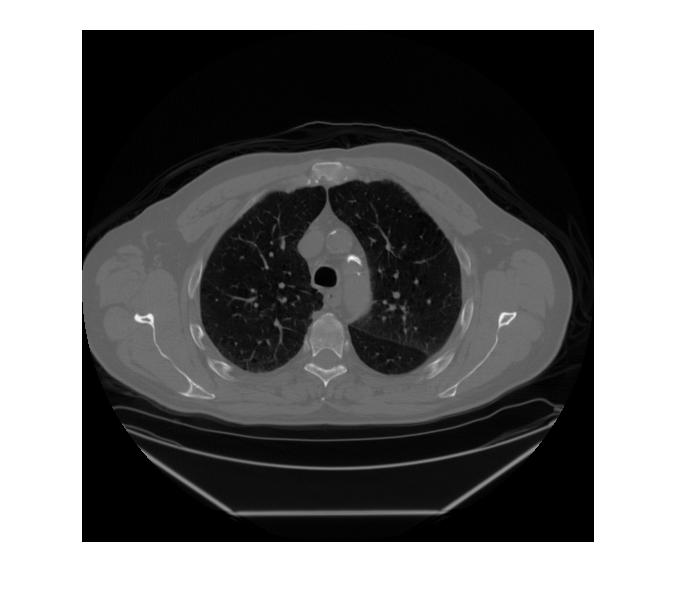}}
\subfigure[]{\includegraphics[width = 1in, height=0.8in]{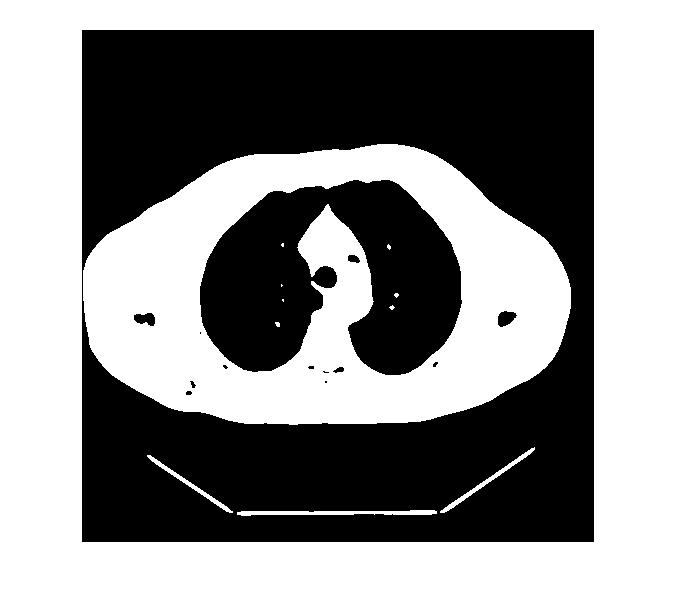}}
\subfigure[]{\includegraphics[width = 1in, height=0.8in]{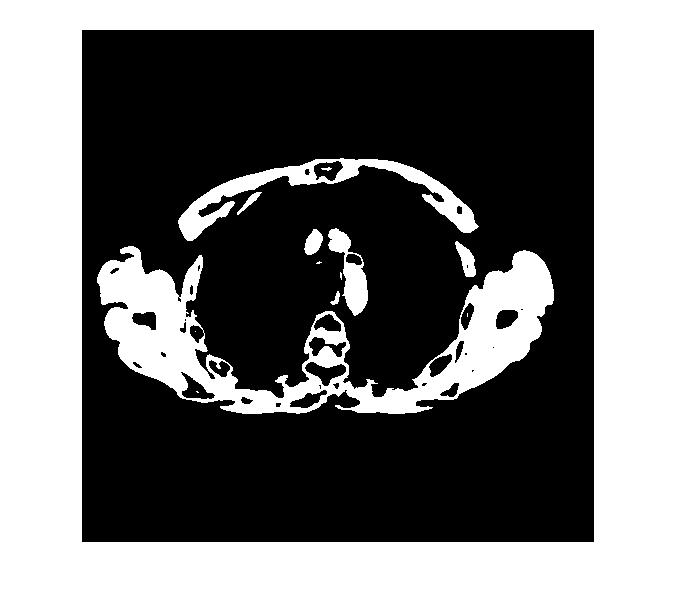}}
\caption{ A CT image from the patient data imaged at 48$mA$ tube current. (a) The original CT image, (b) a mask of ROI acquired by fixed thresholding $[0.5,0.6] = 1$, and (c) a mask of ROI acquired by fixed thresholding $[0.525,0.575] = 1$.}     \label{patroi}
\end{center}
\end{figure}

We begin by visually comparing the ROI selected by the WFDM method to that of the baseline method.
Fig. \ref{patroi} demonstrates the baseline thresholding method from Fig. \ref{roi} now applied to an image taken from a patient at 48$mA$ tube current. Fig. \ref{patroi}(a) is the original image scaled to $[0,1]$ and Fig. \ref{patroi}(b) is the mask acquired by thresholding pixel intensities such that $[0.4,0.7]=1$ and the remaining pixels are set to $0$. The ROI shown in Fig. \ref{patroi}(c) results from further narrowing the threshold so that $[0.5,0.6]=1$ and the remaining pixels are set to $0$. In addition, these threshold ranges were chosen empirically, hence the baseline method is not generalizable.

\begin{figure}[h!]
\begin{center}
\subfigure[]{\includegraphics[width = 1.3in, height=1in]{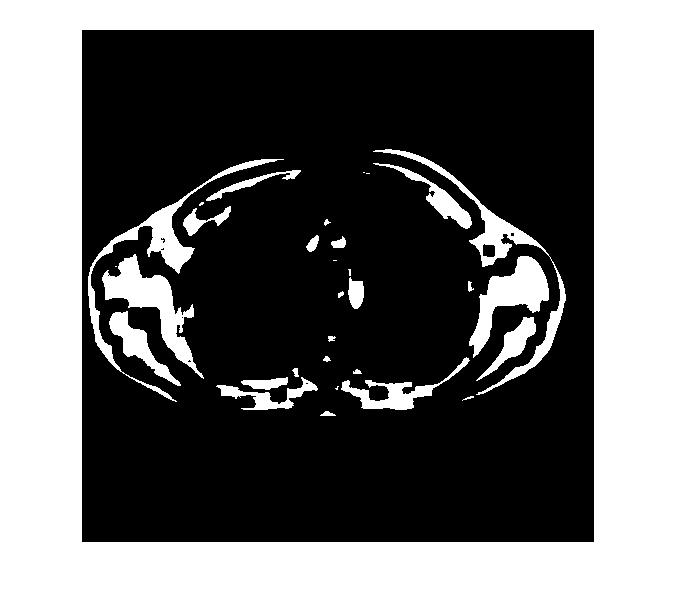}}
\subfigure[]{\includegraphics[width = 1.3in, height=1in]{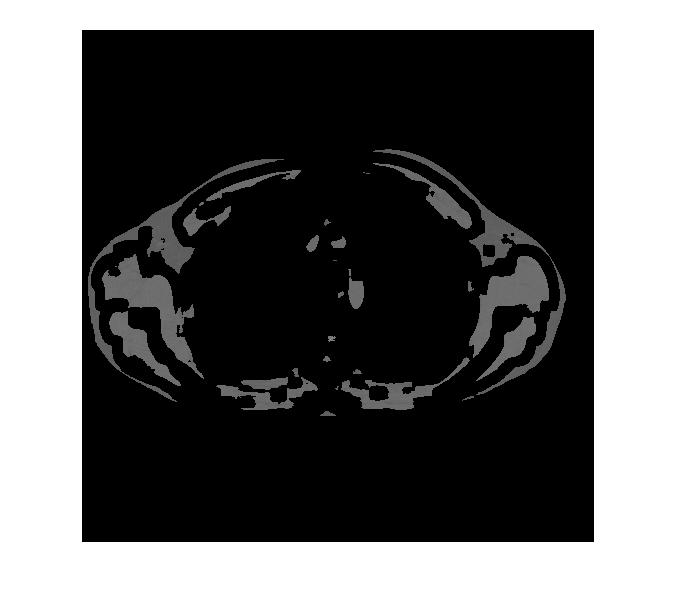}}
\caption{The same CT image as in Fig. \ref{patroi}(a) after processing with WFDM. (a) A mask ROI-LV and (b) the masked ROI-LV are observed.}     \label{patdroi}
\end{center}
\end{figure}

Conversely, applying WFDM to the same CT image shown in Fig. \ref{patroi}(a) and thresholding at $[q<\frac{\overline{\rm I_D}}{4}]$, where $\overline{\rm I_D}$ is the mean distance in $I_D$, results in the mask and corresponding ROI-LV shown in Fig. \ref{patdroi}, respectively. Here, we observe the qualitative variations between the original image and the masked ROI-LV image. First, edges, as well as regions of high spatial variation such as the spine, have been removed. Second, the selected regions are not necessarily of the same absolute pixel intensity, as compared with the ROIs shown in Fig.\ref{patroi}. Finally, the masked ROI-LV region in Fig. \ref{patdroi}(b) is selected by relative thresholding as opposed to fixed intensity thresholding in the baseline method. Thus, the proposed WFDM based ROI-LV masking method is generalizable.

\subsection{Performance Metrics}
The CT images in these datasets are processed by WFDM. The selected ROI-LV are then used to calculate variance, standard deviation, and coefficient of variation of the noisy image, $I_n$, shown in (\ref{var_eq}), (\ref{std}), and (\ref{cov}), respectively. Here, the symbol `$\circ$' represents pixel-wise multiplication.
\begin{align}
\forall U_T^q = 1, \quad
var&: \sigma^2(U^q_T\circ I_n), \label{var_eq}\\
std&: \sigma(U^q_T\circ I_n), \label{std}\\
cov&: \frac{\sigma}{\mu}(U^q_T\circ I_n). \label{cov} 
\end{align}

Since the mask, $U^q_T$, used to extract the ROI-LV regions, depends on the threshold $q$, several values of $q$ are comparatively analyzed. These $q$ threshold values are defined as fractions of the mean of distances in the resulting distance image, $I_D$, for each CT image $I_n$.

The number of pixels below each $q$ threshold value, $\nu$,  is also calculated (\ref{nu}).
\begin{align}
\begin{split}
\nu(q) &= \sum_{x = 1}^n \sum_{y=1}^n (I_D\leq q)(x,y)\\
&= \sum_{x = 1}^n \sum_{y=1}^n U^q_T(x,y).\label{nu}
\end{split}
\end{align}

The abdominal ROI is defined as the tissue region of the CT image, hence the ROI-LV is a subset of the abdominal ROI. Thus, the pixels in the ROI-LV region, $\nu$, are proportional to the number of pixels in the abdominal ROI for each patient CT image. The abdominal ROI is extracted by fixed thresholding, as shown in Fig. \ref{roi}.

Defining the number of pixels in the abdominal ROI as $\nu_M$, the number of pixels in the ROI-LV relative to $\nu_M$ can be calculated as in (\ref{nup}). This fraction of the abdominal ROI in ROI-LV, $\nu'(q)$, accounts for changes in abdominal ROI between patients and is therefore more generalizable.
\begin{align}
\nu'(q) &= \frac{\nu(q)}{\nu_M}. \label{nup}
\end{align}

The fraction of pixels, $\nu'(q)$, is representative of the relatively uniform ROI-LV region. Thus, it can be used to quantitatively analyze the correlation between CT image acquisition quality and the variation in spatial pixel intensities in the ROI-LV region.

\section{Experiments \& Results}

The sensitivity analysis of the metrics $var$ (\ref{var_eq}), $std$ (\ref{std}), $cov$ (\ref{cov}), and $\nu'(q)$ (\ref{nup}) with respect to $
q=[\overline{\rm I_D},\frac{\overline{\rm I_D}}{2},\frac{\overline{\rm I_D}}{3}, \frac{\overline{\rm I_D}}{4}, \frac{\overline{\rm I_D}}{5}, \frac{\overline{\rm I_D}}{6}, \frac{\overline{\rm I_D}}{7}, \frac{\overline{\rm I_D}}{8}, \frac{\overline{\rm I_D}}{9}, \frac{\overline{\rm I_D}}{10}]$
for different tube currents is shown below. Tube currents are directly proportional to CT image noise.

\begin{figure}[h!]
\begin{center}
\subfigure[]{\includegraphics[width = 1.3in, height=1in]{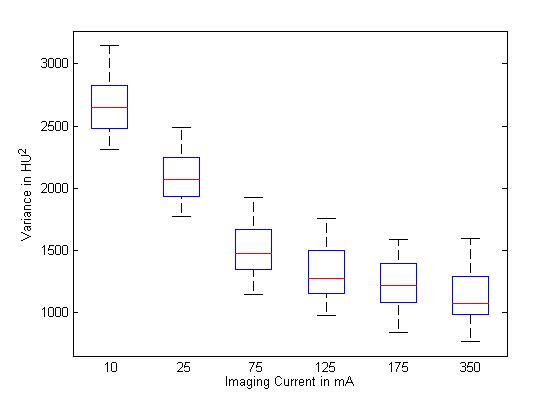}}
\subfigure[]{\includegraphics[width = 1.3in, height=1in]{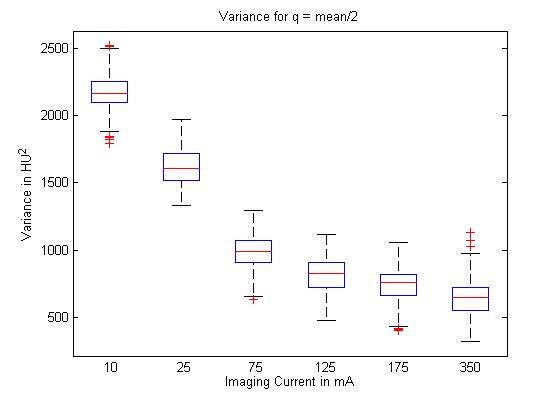}}
\subfigure[]{\includegraphics[width = 1.3in, height=1in]{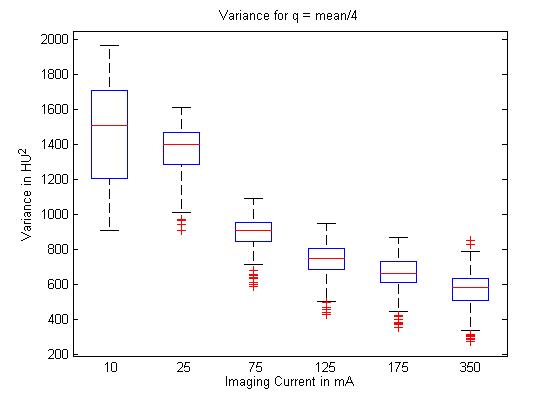}}
\subfigure[]{\includegraphics[width = 1.3in, height=1in]{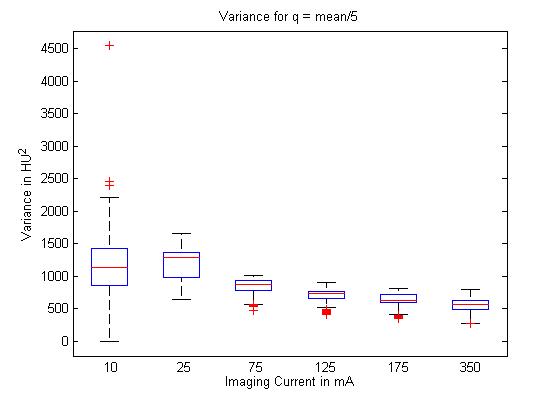}}
\caption{Analysis of $var$ for the phantom CT images. (a) The $var$ within ROI found by fixed thresholding. (b)-(c) The $var$ within ROI-LV found by WFDM where $q = [\frac{\overline{\rm I_D}}{2},\frac{\overline{\rm I_D}}{4},\frac{\overline{\rm I_D}}{5}]$, respectively. The x-axis is the tube current in $mA$. The y-axis is the variance in $HU^2$, where $HU$ is the Hounsfield Unit.}     \label{var}
\end{center}
\end{figure}

For the phantom CT images, the $var$ measured in the baseline ROI is shown in Fig. \ref{var}(a). Here, $var$ is observed to be inversely proportional to the CT image noise. A similar trend is observed for the ROI-LV regions obtained by the proposed WFDM method, such as in Fig. \ref{var}(b), where $q=\frac{\overline{\rm I_D}}{2}$. However, a higher variation in $var$ is observed in the baseline ROI than the ROI-LV. Thus, the ROI-LV of CT images acquired for each tube current are more uniform than the baseline method.

The sensitivity of $var$ to the $q$ threshold value is also observed by a comparative analysis of Fig. \ref{var}(b) - \ref{var}(d). The range of $var$ values is observed to be inversely proportional to CT image noise, and increases with decreasing values of $q$, as shown in Fig. \ref{var}. This observation reflects the high variability with regards to sample size in ROI-LV regions obtained at lower thresholds.

Additionally, the $std$ metric generates similar trends to $var$, with respect to CT image noise. However, the $cov$ results in no significant trends, since $q$ is a function of $\overline{\rm I_D}$.

\begin{figure}[h!]
\begin{center}
\subfigure[]{\includegraphics[width = 0.8in, height=0.8in]{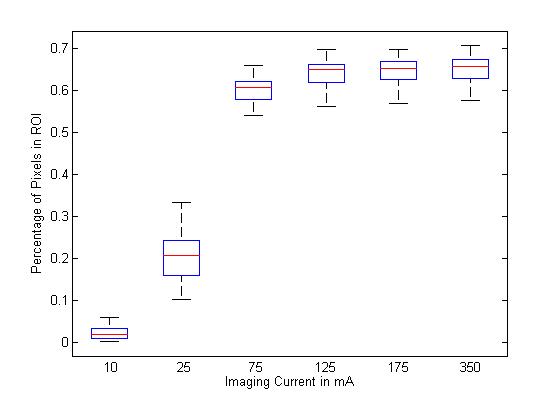}}
\subfigure[]{\includegraphics[width = 0.8in, height=0.8in]{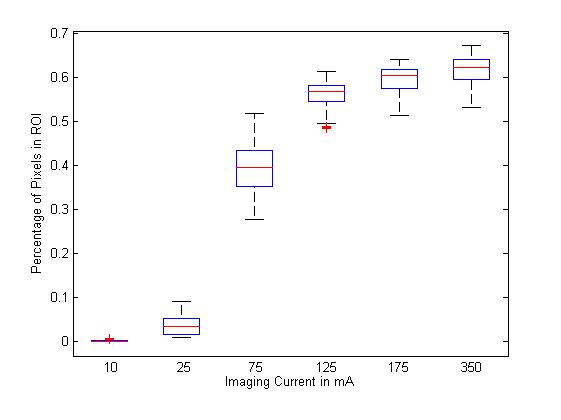}}
\subfigure[]{\includegraphics[width = 0.8in, height=0.8in]{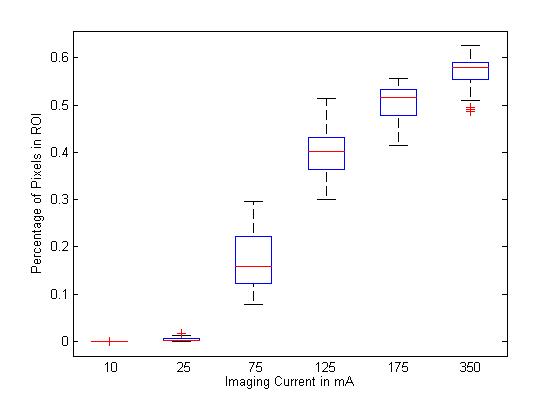}}
\subfigure[]{\includegraphics[width = 0.8in, height=0.8in]{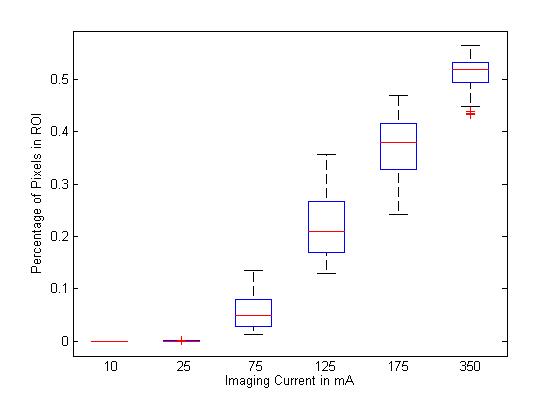}}
\caption{Analysis of $\nu'(q)$ for the phantom CT images. (a) $\nu'(q)$ at $q = \frac{\overline{\rm I_D}}{4}$, (b) $q = \frac{\overline{\rm I_D}}{5}$, (c) $q = \frac{\overline{\rm I_D}}{6}$, and (d) $q = \frac{\overline{\rm I_D}}{7}$, respectively. The x-axis is the tube current in $mA$. The y-axis is the fraction of ROI pixels in ROI-LV.}     \label{box1}
\end{center}
\end{figure}

The sensitivity of the ROI-LV to $q$ is better illustrated by examining the size of the ROI-LV relative to the abdominal ROI, $\nu'(q)$, as shown in Fig. \ref{box1}. Lower quality images show higher sensitivity to $q$ than higher quality images. Thus, the correlation between $\nu'(q)$ and CT image noise for higher quality CT images is increased by decreasing the $q$ threshold value. 

For the phantom CT images, the overall sensitivity analysis for the correlation of each metric, $\nu'(q)$, $var$, $std$, and $cov$, to CT image noise, as a function of the $q$ threshold value, is shown in Table \ref{tab1}. Here, $(*)$ represents $p\ll0.001$.

\begin{center}
\begin{table}[H] 
\centering \caption{Correlation of $\nu'(q)$, $var$, $std$, and $cov$ with tube current in phantom CT images, at different $q$ threshold values.}\label{tab1}
\resizebox{\columnwidth}{!}{
\begin{tabular}{ |c|c|c|c|c| }
\hline
Threshold &\multicolumn{4}{|c|}{Metric}\\
\hline
 &$\nu'(q)$ &$var$ &$std$ &$cov$\\	
\hline
 $q$ &$r\,(p)$ &$r\,(p)$ &$r\,(p)$ &$r\,(p)$\\
\hline
$\overline{\rm I_D}$ &$-0.29\,(*)$&$-0.73\,(*)$&$-0.75\,(*)$&$-0.26\,(*)$\\
\hline
$\frac{\overline{\rm I_D}}{2}$ &$0.21\,(*)$&$-0.74\,(*)$&$-0.76\,(*)$&$-0.28\,(*)$\\
\hline
$\frac{\overline{\rm I_D}}{3}$ &$0.55\,(*)$&$-0.76\,(*)$&$-0.78\,(*)$ &$-0.062\,(0.175)$\\
\hline
$\frac{\overline{\rm I_D}}{4}$ &$0.69\,(*)$&$-0.75\,(*)$&$-0.77\,(*)$ &$0.0044\,(0.924)$\\
\hline
$\frac{\overline{\rm I_D}}{5}$ &$0.79\,(*)$&$-0.53\,(*)$&$-0.49\,(*)$ &$-0.0022\,(0.960)$\\
\hline
$\frac{\overline{\rm I_D}}{6}$ &$0.89\,(*)$&$-0.0010\,(0.826)$&$0.20\,(*)$ &$-0.010\,(0.819)$\\
\hline
$\frac{\overline{\rm I_D}}{7}$ &$0.95\,(*)$&$0.24\,(*)$&$0.41\,(*)$ &$0.020\,(0.656)$\\
\hline
$\frac{\overline{\rm I_D}}{8}$ &$0.96\,(*)$&$0.41\,(*)$&$0.56\,(*) $&$0.053\,(0.241)$\\
\hline
$\frac{\overline{\rm I_D}}{9}$ &$0.95\,(*)$&$0.47\,(*)$&$0.599\,(*)$ &$-0.00037\,(0.993)$\\
\hline
$\frac{\overline{\rm I_D}}{10}$ &$0.91\,(*)$&$0.46\,(*)$&$0.611\,(*)$ &$0.037\,(0.414)$\\
\hline
\end{tabular}}\\

\end{table}
\end{center}

The results in Table \ref{tab1} indicate that the correlation between $\nu'(q)$ and CT image noise is highest at $q = \frac{\overline{\rm I_D}}{8}$, with a correlation coefficient (r-value) of $0.96$ ($p\ll0.001$). On the other hand, both $var$ and $std$ correlate better to CT image noise at $q=\frac{\overline{\rm I_D}}{3}$, with absolute correlation coefficient  $|r|>0.76$ ($p\ll0.001$).

\begin{figure}[h!]
\begin{center}
\subfigure[]{\includegraphics[width = 1.3in, height=1.1in]{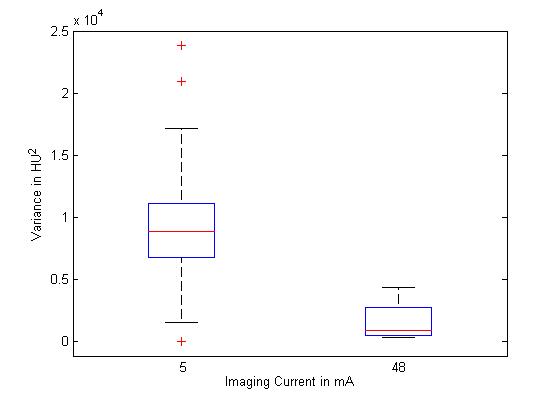}}
\subfigure[]{\includegraphics[width = 1.3in, height=1.1in]{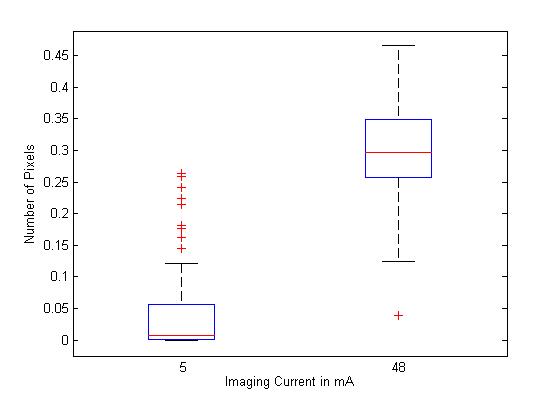}}
\caption{Analysis of (a) $var$ and (b) $\nu'$ for the patient CT images at $q = \frac{\overline{\rm I_D}}{3}$. The x-axis is the tube current in $mA$. The y-axes are (a) variance in $HU^2$, and (b) the fraction of ROI pixels in ROI-LV, respectively. }     \label{pat1}
\end{center}
\end{figure}

Next, the patient CT images are evaluated with respect to the metrics $var$ and $\nu'(q)$, as shown in Fig. \ref{pat1} for $q=\frac{\overline{\rm I_D}}{3}$. Here, the relationship of both $var$ and $\nu'(q)$ to CT image noise reflects the trends observed in the phantom CT images, where $var$ is observed to be inversely proportional to CT image noise while $\nu'(q)$ is observed to be directly proportional to CT image noise. 

\begin{figure}[h!]
	\begin{center}
		\subfigure[]{\includegraphics[width = 1in, height=0.8in]{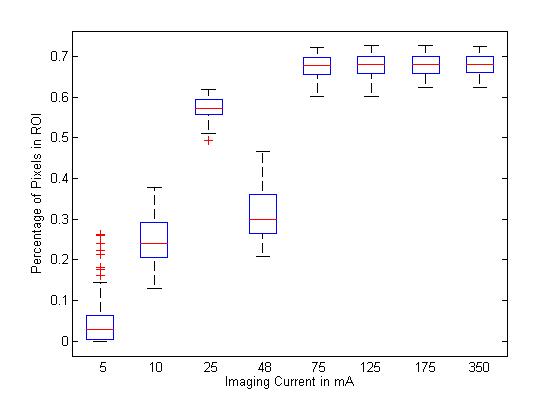}}
		\subfigure[]{\includegraphics[width = 1in, height=0.8in]{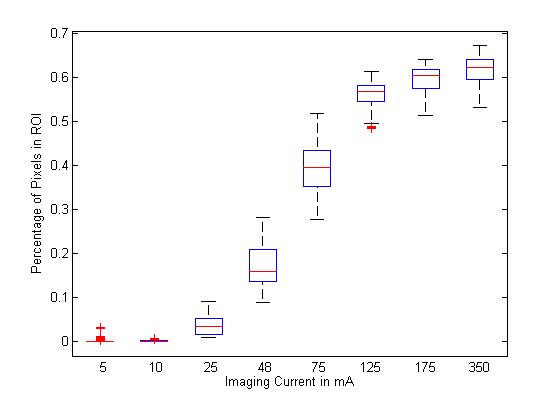}}
		\subfigure[]{\includegraphics[width = 1in, height=0.8in]{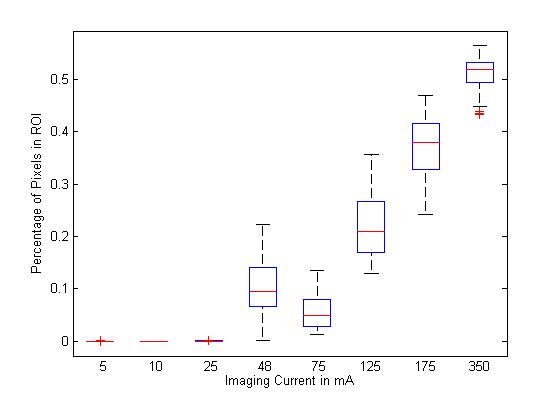}}
		\caption{Analysis of $\nu'(q)$ for phantom CT images at 10, 25, 75, 125, 175, and 350$mA$, and patient CT images at 5 and 48$mA$ respectively. $\nu'(q)$ at (a) $q = \frac{\overline{\rm I_D}}{3}$, (b) $q = \frac{\overline{\rm I_D}}{5}$, and (c) $q = \frac{\overline{\rm I_D}}{7}$, respectively.}     \label{tot1}
	\end{center}
\end{figure}

In addition, $\nu'(q)$ for the combination of patient and phantom CT images at different $q$ threshold values is shown in Fig. \ref{tot1}. Here, the patient CT images are observed to be significantly less sensitive to $q$ than the phantom CT images. This is further supported by the results in Table \ref{tab2}, showing the correlation of each metric $\nu'(q)$, $var$, $std$, and $cov$ to CT image noise at different $q$ threshold values, for the combination of patient and phantom CT images. Here again, $(*)$ represents values of $p<0.001$.
\begin{center}
	\begin{table}[H] 
		\centering \caption{Correlation of $\nu'(q)$, $var$, $std$, and $cov$ with tube current in patient and phantom CT images, at different $q$ threshold values.}\label{tab2}
		\resizebox{\columnwidth}{!}{
		\begin{tabular}{ |c|c|c|c|c| }
			\hline
			Threshold &\multicolumn{4}{|c|}{Metric}\\
			\hline
			&$\nu'(q)$ &$var$ &$std$ &$cov$\\	
			\hline
			$q$ &$r\,(p)$ &$r\,(p)$ &$r\,(p)$ &$r\,(p)$\\
			\hline
			$\overline{\rm I_D}$ &$0.23\,(*)$&$-0.38\,(*)$&$-0.46\,(*)$&$0.021\,(0.601)$\\
			\hline
			$\frac{\overline{\rm I_D}}{2}$&$0.40\,(*)$&$-0.32\,(*)$&$-0.45\,(*)$&$0.013\,(0.737)$\\
			\hline
			$\frac{\overline{\rm I_D}}{3}$ &$0.62\,(*)$&$-0.31\,(*)$&$-0.43\,(*)$ &$0.0040\,(0.919)$\\
			\hline
			$\frac{\overline{\rm I_D}}{4}$ &$0.75\,(*)$&$-0.23\,(*)$&$-0.34\,(*)$ &$0.021\,(0.600)$\\
			\hline
			$\frac{\overline{\rm I_D}}{5}$ &$0.83\,(*)$&$-0.14\,(*)$&$-0.21\,(*)$ &$0.0042\,(0.916)$\\
			\hline
			$\frac{\overline{\rm I_D}}{6}$ &$0.90\,(*)$&$-0.12\,(0.002)$&$-0.0057\,(0.886)$ &$-0.017\,(0.670)$\\
			\hline
			$\frac{\overline{\rm I_D}}{7}$ &$0.95\,(*)$&$-0.082\,(0.036)$&$0.084\,(0.033)$ &$0.020\,(0.610)$\\
			\hline
			$\frac{\overline{\rm I_D}}{8}$ &$0.95\,(*)$&$-0.058\,(0.138)$&$0.14\,(*)$ &$0.046\,(0.245)$\\
			\hline
			$\frac{\overline{\rm I_D}}{9}$ &$0.92\,(*)$&$-0.051\,(0.193)$&$0.16\,(*)$ &$0.0050\,(0.900)$\\
			\hline
			$\frac{\overline{\rm I_D}}{10}$ &$0.86\,(*)$&$-0.052\,(0.189)$&$0.16\,(*)$ &$0.022\,(0.570)$\\
			\hline
			
		\end{tabular}}\\
		
	\end{table}
\end{center}

Table \ref{tab2} shows that $\nu'(q)$ is best correlated to CT image noise in both patient and phantom CT images at $q=\frac{\overline{\rm I_D}}{7}$ and $q=\frac{\overline{\rm I_D}}{8}$, with a correlation coefficient of $0.95$ ($p\ll0.001$). On the other hand, the $var$, $std$, and $cov$ do not correlate well to CT image noise in the combined patient and phantom CT images, with absolute correlation coefficients below 0.46 ($p\ll0.001$).

Although the trends observed in Fig. \ref{var}, \ref{box1}, and \ref{pat1}, demonstrate that both $var$ and $\nu'(q)$ correlate with CT image noise, the absolute values of $var$ between the phantom and patient CT image datasets differ by an order of magnitude, whereas the value of $\nu'(q)$ is relative to the size of the abdominal ROI in each image. Thus, $var$ does not correlate as well to CT image noise in the combined phantom and patient CT images as $\nu'(q)$.

Finally, the sensitivity of $\nu'(q)$ to $q$ decreases at lower $q$ threshold values, as observed in Table \ref{tab1} and Table \ref{tab2}. In addition, lower values of $q$ result in larger variations in size of ROI-LV within CT images acquired at each tube current, as observed in Fig. \ref{box1} and Fig. \ref{tot1}. As a result, $q$ can be optimized relative to both.

\section{Conclusion \& Discussion}
In this paper, a novel WFDM approach is presented for CT image noise detection and evaluated for both phantom and patient CT images using $var$, $std$, $cov$, and $\nu'(q)$ as metrics, given that CT image noise is directly proportional to tube current. For the phantom CT images,  a correlation between both $var$ and $std$ to CT image noise ($r>0.76$ ($p\ll0.001$)) and $\nu'(q)$ to CT image noise ($r=0.96$ ($p\ll0.001$)) is observed.

However, after combining the phantom and patient CT images, $var$ and $std$ do not correlate well to CT image noise($r<0.46$ ($p\ll0.001$)) as compared to $\nu'(q)$ ($r=0.95$ ($p\ll0.001$)). This suggests that $\nu'(q)$ is a more generalizable metric for CT image noise detection than $var$ and $std$.

Future work will be directed towards the analysis of $\nu'(q)$ with additional patient datasets to find an optimal choice of $q$, and further evaluate $\nu'(q)$ as a metric for CT image noise. The ROI-LV will also be evaluated with other variance based metrics for comparison. Finally, the noise in this experiment is assumed to be normally distributed. Therefore, the relationship between different types of noise and the WFDM will be evaluated to assess the impact of this assumption on the resulting CT IQ classification.

\bibliographystyle{IEEEtran}
\bibliography{IEEEabrv,bibtest3}
\end{document}